# PSF-based Analysis for Detecting Unresolved Wide Binaries

You Wu[1], Jiao Li[1], Chao Liu[1], Yi Hu[2], Long Xu[3], Tanda Li[4], Xuefei Chen[5,6], and Zhanwen Han[5,6]
[1] Key Laboratory of Space Astronomy and Technology, National Astronomical Observatories, Chinese Academy of Sciences, Beijing 100101, People's Republic of China; lijiao@bao.ac.cn, liuchao@nao.cas.cn
[2] National Astronomical Observatories, Chinese Academy of Sciences, Beijing 100101, People's Republic of China
[3] Key Laboratory of Solar Activity, National Astronomical Observatories, Chinese Academy of Sciences, Beijing 100101, People's Republic of China; lxu@nao.cas.cn
[4] Department of Astronomy, Beijing Normal University, Beijing 100875, People's Republic of China
[5] Yunnan Observatories, CAS, P.O. Box 110, Kunming 650011, Yunnan, People's Republic of China
[6] Center for Astronomical Mega-Science, Chinese Academy of Science, Beijing 100012, People's Republic of China
Received 2022 September 1; revised 2023 July 7; accepted 2023 July 20; published 2023 September 11

## Abstract

Wide binaries play a crucial role in analyzing the birth environment of stars and the dynamical evolution of clusters. When wide binaries are located at greater distances, their companions may overlap in the observed images, becoming indistinguishable and resulting in unresolved wide binaries, which are difficult to detect using traditional methods. Utilizing deep learning, we present a method to identify unresolved wide binaries by analyzing the point-spread function (PSF) morphology of telescopes. Our trained model demonstrates exceptional performance in differentiating between single stars and unresolved binaries with separations ranging from 0.1 to 2 physical pixels, where the PSF FWHM is ∼2 pixels, achieving an accuracy of 97.2% for simulated data from the Chinese Space Station Telescope. We subsequently tested our method on photometric data of NGC 6121 observed by the Hubble Space Telescope. The trained model attained an accuracy of 96.5% and identified 18 wide binary candidates with separations between 7 and 140 au. The majority of these wide binary candidates are situated outside the core radius of NGC 6121, suggesting that they are likely first-generation stars, which is in general agreement with the results of Monte Carlo simulations. Our PSF-based method shows great promise in detecting unresolved wide binaries and is well suited for observations from space-based telescopes with stable PSF. In the future, we aim to apply our PSF-based method to next-generation surveys such as the China Space Station Optical Survey, where a larger-field-of-view telescope will be capable of identifying a greater number of such wide binaries.

*Unified Astronomy Thesaurus concepts:* Binary stars (154); Globular star clusters (656); Convolutional neural networks (1938); Photometry (1234)

## 1. Introduction

Binary systems are a ubiquitous product of the star formation process. Roughly half of all main-sequence (MS) stars exist in multiple systems, in which two or more stellar components are commonly involved (Duquennoy & Mayor 1991; Raghavan et al. 2010; Moe & Di Stefano 2017). The orbital separation of these systems is distributed over a wide range (Duquennoy & Mayor 1991; Duchêne & Kraus 2013). A complete statistical analysis for companions to solar-type stars was presented by Raghavan et al. (2010), which revealed that almost half of solar-type stars within 25 pc of the Sun have companions and the median of period distribution is ∼300 yr. Tokovinin & Lépine (2012) investigated the multiplicity of solar-type dwarfs within 67 pc of the Sun and found that the proportion of binaries with a separation between 2 and 64 kau is higher than 4.4%.

Wide binaries, being weakly gravitationally bound, provide a sensitive probe for astrophysics to study the Galactic dynamical evolution (e.g., Weinberg et al. 1987; Jiang & Tremaine 2010; Deacon & Kraus 2020). Since the components of wide binaries are not expected to interact with each other during their life, they can be thought to have evolved in isolation, and tracing their origins can contribute to our understanding of the environment in which stars were born (Retterer & King 1982; Sterzik et al. 2003). The current mechanisms of wide binary formation support the idea that the components of wide binaries are essentially coeval with similar chemical abundances (e.g., Kouwenhoven et al. 2010; Moeckel & Clarke 2011; Tokovinin 2017; Hawkins et al. 2020; Nelson et al. 2021), which allows them to contain information that places potent constraints on stellar physics (Andrews et al. 2018). For example, wide binaries have been utilized to calibrate metallicity indicators (e.g., Bonfils et al. 2005; Rojas-Ayala et al. 2010) and age–rotation relations (e.g., Barnes 2007; Chanamé & Ramírez 2012; Godoy-Rivera & Chanamé 2018) and to constrain the initial-to-final mass relation for white dwarfs (e.g., Catalán et al. 2008; Andrews et al. 2015; Barrientos & Chanamé 2021). In addition, the properties of wide binaries in clusters are decisive for investigating the initial density and dynamical processes of clusters (e.g., Scally et al. 1999; Parker et al. 2009).

Currently, several primary techniques are used to identify binaries. The first method is by measuring their radial velocity variability (e.g., Duquennoy & Mayor 1991; Nidever et al. 2002; Geller et al. 2015), which is affected by the eccentricity of the orbit and is only applicable to short-period binaries. The second method derives from the photometric variables (e.g., Carney 1983; Strassmeier et al. 1989; Prša et al. 2011). One can deduce whether two stars orbit each other in a tight orbit by observing how light varies periodically with time, but this method is affected by the orbital inclination and tends to find binaries with short periods. The third approach is feasible only for binaries with the largest separations, i.e., by measuring their common proper motion across the sky, provided that the components can be resolved and distinguished in the observational data. (e.g., Luyten 1971; Wasserman & Weinberg 1991; Chanamé & Gould 2004; Oh et al. 2017; El-Badry & Rix 2018; Hartman & Lépine 2020). The fourth method, employing the astrometric technique, involves accurately measuring the positions and motions of stars in the sky. In instances where an unseen companion star is present, the observable star exhibits a periodic wobble in its motion, referred to as the astrometric signature. By analyzing this signature, we can deduce the existence of the companion star and determine the properties of the binary system (Halbwachs et al. 2023; Penoyre et al. 2022a, 2022b). In particular, for clusters, there is a statistical method to investigate the populations of binaries by analyzing binary sequences located on the red side of the single MS (e.g., Romani & Weinberg 1991; Bellazzini et al. 2002; Richer et al. 2004; Sollima et al. 2007a; Milone et al. 2009). This method is independent of orbital period and inclination but is affected by photometric accuracy (Milone et al. 2012).

For wide binaries, neither the variables of radial velocity nor photometric variables can be easily observed owing to their long period ($P_{orb} \sim 10^2$–$10^6$ yr or more). In situations where wide binaries are located at significant distances from us, the components might overlap in the observed images, resulting in the appearance of a single point source, despite the fact that they are considerably separated from each other.

Consider, for instance, a wide binary comprising two equal-mass components of 1 $M_\odot$ each, with a 100 yr orbital period and a semimajor axis of 27.14 au. If this binary were located at a distance of 1 kpc, it would not be spatially resolved as two isolated points by a telescope with a spatial resolution limit of $R > 27$ mas. This limitation arises from the spatial resolution capabilities of telescopes, which are fundamentally determined by diffraction effects and constrain the ability to distinguish two close objects. When the angular separation between two stars in a binary is smaller than the spatial resolution limit of a telescope, the light from each star is combined, resulting in the binary appearing as an unresolved point source in astronomical observations. Therefore, it is impossible to identify such wide binaries by proper motions and astrometric parallaxes. Several methods are available to identify unresolved binaries, e.g., the direct spectral detection method (Gullikson et al. 2016), diffraction-limited imaging (Hubrig et al. 2001), and a Bayesian model in color–magnitude space (Widmark et al. 2018), but they are based on, e.g., spectral analysis, searching for specific types of systems, or analysis of the color–magnitude diagram (CMD), none of which is performed directly on photometric images. Furthermore, these techniques may require multiple observations and extensive telescope time to achieve. The development of appropriate approaches to detect unresolved binaries is still a challenge that remains to be addressed. Indeed, determining whether a point source in the photometric image is a single star or not carries significant importance, as this enables us to better understand the multiplicity of stars and assign additional observation constraints on the binary distributions.

With the advent of deep learning, a data-driven approach to image analysis has been implemented that relies not on manual extraction of features but on automatic learning of deep abstract features by machines (LeCun et al. 2015; Goodfellow et al. 2016). Deep learning has been proven to be a significant success in computer vision tasks such as identification, generation, and classification (Voulodimos et al. 2018). It is also proving to be outstanding in many areas of astronomy. For example, deep learning has been applied to searching special objects from spectral surveys (e.g., Parks et al. 2018; Shallue & Vanderburg 2018), determining the stellar parameters (e.g., Fabbro et al. 2018; Leung & Bovy 2019), detecting extreme-ultraviolet waves from solar bursts (e.g., Xu et al. 2020), and analyzing the morphological structures of galaxies (e.g., Domínguez Sánchez et al. 2018; Barchi et al. 2020). Considering the advantages of deep learning in representative feature extraction, this could be a promising tool to provide a solution to study unresolved pairs by further learning multiple levels of data representations.

Another crucial consideration in the analysis of unresolved binaries is how to obtain high-quality images. During the process of imaging astronomical data, there are varying degrees of distortion owing to the effects of the point-spread function (PSF), photoelectric noise, background noise, etc. (Starck & Murtagh 2013). The PSF dominates the morphology of astronomical images and is generally influenced by factors such as aberration, diffraction limit, and atmospheric disturbances (Racine 1996). Ground-based telescope imaging is most affected by the atmospheric disturbance that adds stochasticity to the PSF (Perrin et al. 2003). Although adaptive optics systems can help with wave front correction (Beckers 1993), the loss of detail due to image degradation cannot be entirely avoided in view of the limitations and complexity of the system. (Davies & Kasper 2012). Space-based telescopes, on the contrary, are deployed above the atmosphere to avoid atmospheric disturbances, and their PSF is mainly influenced by the diffraction limit of the instrument itself, resulting in high photometric accuracy and image quality with a good PSF shape. To specifically distinguish between unresolved pairs and individual luminous stars, a sufficient number of high-quality images are required. Such photometric data are available from large surveys of space-based observatories with high spatial resolution imaging.

One prospective mission that will achieve this goal will be the Chinese Space Station Telescope (CSST; Zhan 2011; Cao et al. 2018). CSST is a space telescope with an aperture of 2 m, scheduled for launch in 2024. It is capable of conducting photometric surveys and slitless grating spectroscopic surveys over an area of 17,500 $deg^2$ of sky, with wavelength coverage ranging from near-ultraviolet to near-infrared (255–1000 nm). CSST has a large field of view of $\sim$1 $deg^2$ with a high spatial resolution of $\sim$0.″15 (80% energy concentration region). The survey instruments will obtain billions of photometric data of stars and galaxies, as well as hundreds of millions of spectra over a 10 yr period. We therefore expect that an abundant sample of wide binaries would be identified from the high spatial resolution images by performing an analysis of CSST data.

The primary goal of this study is to develop a valuable and computationally efficient method for detecting unresolved wide binaries. We present the first application of a deep-learning approach for this purpose, analyzing photometric images from space-based observatories. This work serves as a preparatory step toward future programs with the CSST. In this regard, we conduct experiments on simulated data from CSST and assess the feasibility of the proposed method using real observational data obtained from the Hubble Space Telescope (HST). Our analysis in this work involves the robustness and limitations of unresolved binaries detection, which also prepares for the next generation of surveys.

The remainder of this paper is structured as follows. In Section 2, we outline the methods used for detecting wide binaries, including the construction of the training set, as well as the implementation and optimization of the deep-learning model. Section 3 presents the performance of our model for CSST and details the experiments of the factors that affect the model. We validate our approach in HST data and present a discussion in Section 4, in which we describe how to generate simulated HST data with the PSF and show the results of a search for wide binaries in NGC 6121 using our method. Finally, in Section 5, we conclude with a brief summary, followed by a discussion of future prospects.

## 2. Method

### 2.1. Contamination of Chance Alignments

Before discussing the detection of wide binaries, it is necessary to assess the contamination from chance alignments that affects the feasibility essentially. Chance alignments are not physically bound binaries, whose effects have been considered in many investigations searching for samples of wide binaries (e.g., Hawkins et al. 2020; Tian et al. 2020; El-Badry et al. 2021). El-Badry et al. (2021) used two methods to estimate the contamination of chance alignments and show that their contribution is dominant when binary separations are greater than 30,000 au. Despite that these estimations do not take into account the case of chance alignments with closer angular separations, it is apparent that the contamination rate increases with increasing binary separations (see Figure 3 of El-Badry et al. 2021).

In our study, chance alignments refer to unresolved stars in which two stars happen to be in the same line of sight and overlap each other in the image. To statistically estimate their contamination, we designed an experiment based on the observational characteristics of CSST. We generated varying numbers of single stars at regular intervals in a 1 deg$^2$ patch of the sky (equivalent to CSST's field of view) using Monte Carlo simulations, ranging from $10^6$ to $10^7$ stars. We assumed that any two stars within 0″15 are chance alignments, which is in accordance with the spatial resolution of CSST. In other words, CSST can only distinguish two stars if their angular separation is greater than 0″15. This conservative assumption estimates the upper limit of chance alignment. The reason for making this assumption is that our ultimate goal is to detect unresolved binary stars utilizing CSST, so we need to take into account the effect of chance alignments among these unresolved binary stars. We defined the contamination rate as the number of chance alignments divided by the total number of stars. We performed 10 experiments for each stellar density and took the average as the contamination rate, which was plotted in Figure 1, along with the standard deviation as the error bars.

Clearly, the contamination rate demonstrates a positive correlation with the increase in stellar density, exhibiting a linear relationship. It is important to note that the error bars for the contamination rate are quite small, indicating a high level of confidence in the observed trend. At a stellar density of

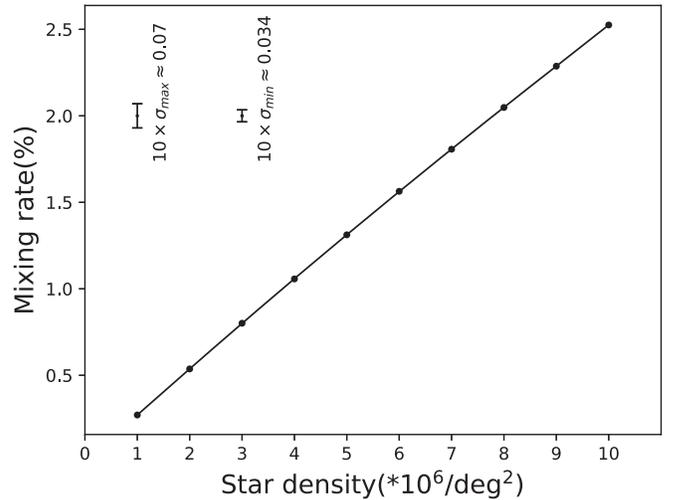

**Figure 1.** Comparison of contamination rate of chance alignments and total number of stars for 1 deg$^2$ of sky. The black circles represent the contamination rates at each stellar density. The line plot formed by connecting these circles with a black line highlights the trend of the data. The error bars in the upper left region of the figure represent the maximum and minimum standard deviations of the contamination rates.

10 million stars per square degree of sky, the contamination rate reaches approximately 2.5%. Meanwhile, for the stellar density $<10^6$ deg$^{-2}$, the contamination rate of chance alignments remains below 0.3%.[7] This trend indicates that a higher stellar density can affect the accuracy of identifying physically bound binaries to a certain extent, while the contamination from chance alignments remains relatively insignificant at lower stellar densities. Consequently, it is crucial to account for the contamination rates arising from chance alignments when analyzing binaries in regions of varying stellar densities.

### 2.2. Data Set Generation for CSST

Obtaining a sufficient number of high-quality training samples is crucial for the successful implementation of deep-learning approaches. The CSST has a multicolor imaging capability covering a wavelength range from 0.255 to 1.0 μm, consisting of eight bands in total. Generally, shorter wavelengths yield higher spatial resolutions of a telescope, while longer wavelengths result in lower resolutions. As our method is based on the analysis of one-band images and this study is primarily aimed at validation, we chose to utilize the PSF of $u$ band (effective wavelength approximately 463.1 nm) for generating mock point-source images in our experiments. The FWHM of the PSF for the $u$ band in the CSST is approximately 0″15. The original CSST PSF is represented by a 256 × 256 matrix (see Figure 2(a)). To generate higher-quality training samples, we subdivided each pixel of the original CSST PSF into smaller subpixels, resulting in a PSF model with a more detailed structure. Convolution operations are then applied to the PSF model, thereby enabling the simulation of point-source imaging at arbitrary locations, including subpixel locations. The resulting virtual point-source images exhibited increased

---

[7] In order to provide a reference for the scale of stellar density in different regions of the Milky Way, we used Gaia DR3 data to roughly calculate the stellar density per square degree of sky in various locations across the Milky Way disk and halo. Based on these estimates, the stellar density in the vicinity of the Milky Way disk is approximately $10^5$ deg$^{-2}$, whereas in the high galactic latitude regions of the halo the stellar density is less than $10^4$ deg$^{-2}$.

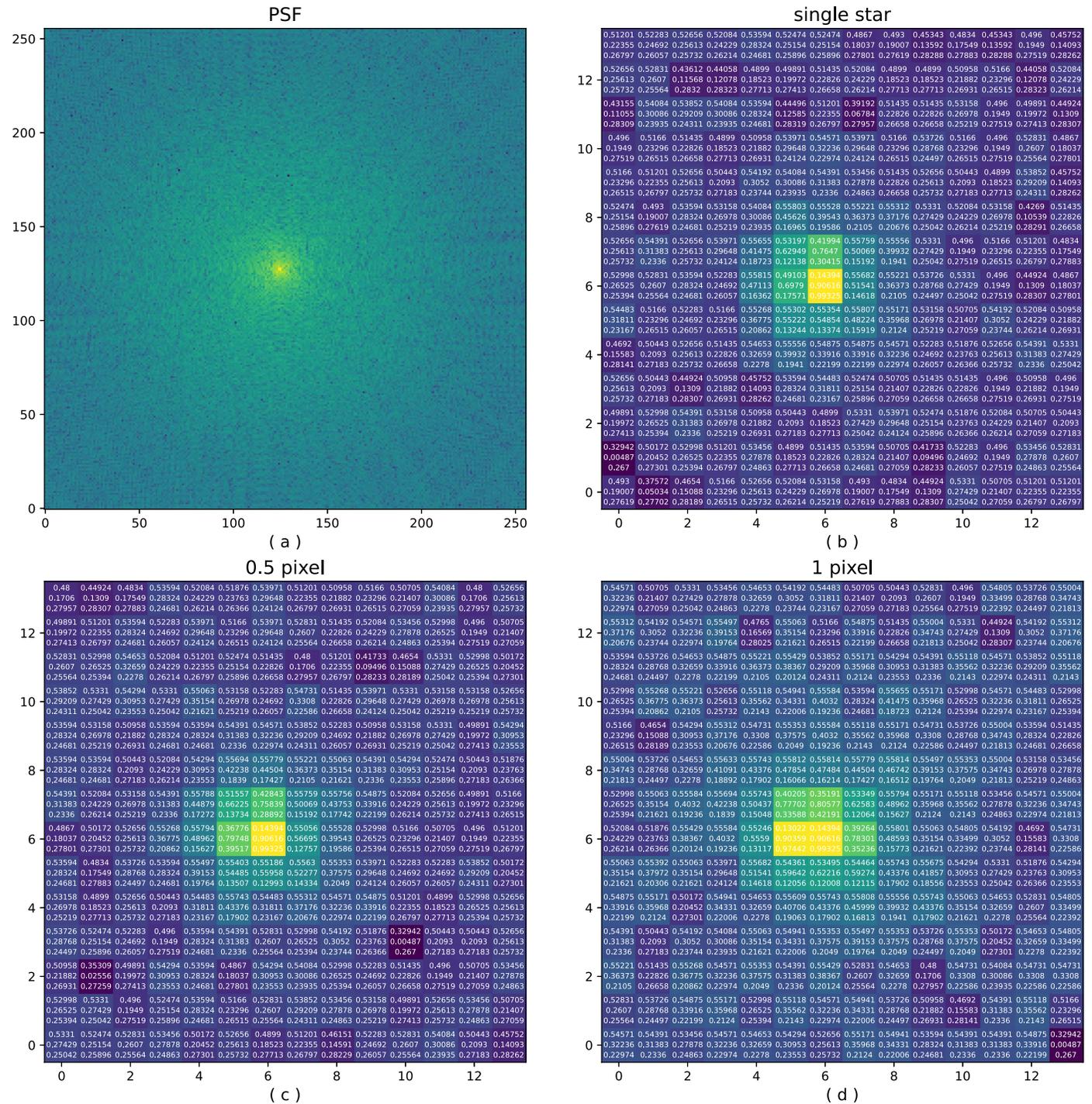

**Figure 2.** Examples of the PSF and simulated images for CSST. (a) Log image of the original PSF for the wavelength of 463.1 nm. (b) Log image of a single star and its corresponding RGB value matrix (flux of star: $F = 10{,}000$). (c) Log image of a binary with 0.5 physical pixel separation and its corresponding RGB value matrix (fluxes of the two stars: $F_1 = F_2 = 5000$). (d) Log image of a binary with 1 physical pixel separation and its corresponding RGB value matrix (fluxes of the two stars: $F_1 = F_2 = 5000$).

fidelity, which can be used as high-quality training samples for deep learning.

Our data sets contain two classes, namely, single stars and binaries. For the single-star class, we generated each mock image by convolving the PSF model with a point source, which is positioned at diverse locations throughout the image to ensure that the generated point sources can appear anywhere, not exclusively at the center. Each point source is assigned a flux, a variable that governs the signal-to-noise ratio (S/N). To prevent introducing implicit bias into the model during the training process, the fluxes[8] of single stars follow a uniform distribution in the range of 1000 to 25,000. In terms of background estimation, some studies, such as the one conducted by He et al. (2021), utilize generative adversarial networks to estimate backgrounds in wide-field observational images. However, our methodology differs owing to our focus on individual sources within relatively smaller regions, which significantly simplifies the task of background estimation. Specifically, for our CSST mock images, we include both Gaussian noise and Poisson noise as part of the backgrounds in the generation process. This approach is intended to better approximate the conditions of real-world observations, enhancing both the robustness and the generalizability of our model and facilitating a more convenient analysis of the S/N. Here the S/N is defined as

$$\mathrm{S/N}(F, \sigma) = \frac{\sum_i^n (F_i + \eta_i)}{\sqrt{\sum_i^n F_i + \sigma^2}}, \eta \sim N(0, \sigma^2), \quad (1)$$

where $F_i$ is the photon flux of pixels and $\eta_i$ is the noise of pixel $i$ that obeys a Gaussian distribution with a mean of 0 and standard deviation of $\sigma$. In Section 3.2, we conduct an experiment to compare the impact of varying S/N on model performance.

---

[8] Please note that the flux mentioned in this paper refers to the number of photons.

For the class of binaries, two point sources are generated in each image, with each point source being assigned a flux value and a set of coordinates, following a process similar to the one described for single stars. By freely adjusting the positions of each point source, we can obtain simulated images of two point sources' various separations. Due to the pixel size and spatial resolution of the photometric instrument in CSST, approximately 0″.075 and 0″.15, respectively, stars within 2 pixels[9] of each other are unresolvable. Furthermore, Hu et al. (2011) employed the artificial-star test technique, demonstrating that a separation of 2 pixels between stars is the minimum distance required for them to be detected as separate objects by the photometry. Consequently, we set the maximum separation for binary stars at 2 pixels and the minimum separation at 0.1 pixels.[10] The minimum separation is chosen to account for the significant uncertainty in detecting binaries at such close proximity.

To eliminate the model bias caused by an uneven distribution of binary separations, which can result in a model being biased toward predicting more frequent values and performing poorly for less frequent values during training, we balanced the binary separation to be uniformly distributed over a range from 0.1 to 2 physical pixels in steps of 0.1 pixels. This ensured that the model was exposed to a variety of binaries with a diverse range of separations, enabling it to effectively learn their characteristics. Additionally, we have carefully controlled the total flux of binaries to be uniformly distributed within the same range as that of the single-star class, which is set between 1000 and 25,000. This minimizes the impact of flux while maintaining consistency between the single-star and binary-star classes. By doing so, we are able to provide a fair evaluation of our model's performance on both classes, without any undue influence from flux variations.

For all mock images, we only retained the central window of $14 \times 14$ pixel areas for two reasons. The profiles of the stars are mainly concentrated on a few pixels. In addition, due to the smaller image size, the computation speed of the training process can be accelerated.

Finally, we generated about 72,840 mock images for CSST, with half being single stars and the other half binaries. The choice of 72,840 images balances the need for a sufficiently large data set to train our model while maintaining a manageable computational workload. Figure 2 illustrates the original PSF of CSST (panel (a)) and shows three examples of mock images from left to right along with their corresponding RGB value matrices, namely, a single star (panel (b)), a binary with 0.5-pixel separation (panel (c)), a binary with 1-pixel separation (panel (d)). Please note that the background level and total flux are the same in these three examples. These images are derived from single-band observations, and while they do not contain color information, their representation in RGB format helps to highlight intensity differences, thereby enhancing feature recognition. As displayed in Figure 2, the stars appear to be compact owing to the concentration of the PSF energy in a small region at the center. By examining the RGB value matrices, we can observe the subtle changes in values between the different cases. For instance, in the case of a binary with 0.5-pixel separation (panel (c)), the centers of the two stars are situated halfway between adjacent pixels, resulting in the values of the central region being affected by the combined flux of both stars. In the case of a binary with 1-pixel separation (panel (d)), the influence of each star is more distinct. These subtle differences in the RGB value matrices, although not easily discernible to the human eye, provide the foundation for a deep-learning approach to extract information hidden in noisy images and further distinguish between single stars and binaries.

### 2.3. Data Augmentation and Preprocessing

Data augmentation is currently the most effective preprocessing technique applied to deep-learning models (Perez & Wang 2017). It is a regularization technique that performs various kinds of transformations on the image, such as geometric transformations, color space transformations, and generative modeling augmentation (Shorten & Khoshgoftaar 2019). Appropriate image transformation can greatly improve the generalization of the model and help introduce more diversity into the training set (Wong et al. 2016; Mikołajczyk & Grochowski 2018). Given that, we adopted the following data augmentation techniques:

1. `Flip`. Randomly reverse the rows or columns of pixels horizontally or vertically.
2. `Rotation`. Apply a random rotation to each image by up to 30°.
3. `Zoom`. Randomly zoom each image with multiples between 1.0 and 1.5.
4. `Warp`. Randomly change the perspective at which the image is viewed.
5. `Brightness` and `Contrast`. Randomly change the amount of light and the contrast of each image.

---

[9] In our study, the pixels we refer to are physical pixels, which are the smallest discrete elements of the imaging detector used to measure separation. The physical pixel is designed to match the spatial resolution of the imaging system, but it is important to note that it does not determine this resolution itself. The spatial resolution of a space telescope is fundamentally dictated by the diffraction limit, which is determined by the diameter of the mirror and the wavelength of the observed light.

[10] The separation referred to here is the distance between the centers of two stars after imaging. Within a spatial region represented by 1 pixel, multiple objects or structures can be contained. For CSST, 1 pixel is 0″.075 in size. Therefore, when the distance between two stars is less than 1 pixel, e.g., 0.1 pixel, it implies that the angular separation of the two stars is 0″.0075. Consequently, the imaging system captures both stars within the same pixel, resulting in a blending of image information that makes it arduous to distinguish between them.

Data augmentation is a technique that, while not altering the number of samples in the original data set, presents the model with various modified versions of the training data during each epoch. In our case, the original data set consists of the 72,840 mock images generated using the CSST *u*-band PSF before the application of data augmentation. This approach is commonly employed during the training phase to enhance the model's performance by broadening the range of input data variations. To further improve the performance, we extended the application of data augmentation techniques to the inference process, adopting the test time augmentation (TTA) technique (Ayhan & Berens 2018; Radosavovic et al. 2018). TTA facilitates accurate image classification by generating multiple augmented variants of each test image and subsequently aggregating the predictions to produce a final output. This process results in smoother and more robust predictions without requiring additional model training.

### 2.4. Network Architecture

A convolutional neural network (CNN) is a multilayer neural network designed using the Back-Propagation algorithm (LeCun et al. 1989; Lecun et al. 1998). The basic functional structure of CNN typically consists of three parts, namely the convolution layers, the pooling layers, and the fully connected layers (O'Shea & Nash 2015). It is a widely used architecture in many computer vision tasks since the convolution kernels are well adapted for feature extraction and significantly reduce the complexity of the model (Razavian et al. 2014). The network is generally extended by adding the number of layers or increasing the width of the network to achieve better performance, as in some popular CNN architectures, e.g., AlexNet (Krizhevsky et al. 2012), VGGNet (Simonyan & Zisserman 2014), GoogLeNet (Szegedy et al. 2015), and ResNet (He et al. 2016). In addition, the larger image resolution can help improve the accuracy as well (Huang et al. 2019).

Balancing the scaling of network width, depth, and resolution is essential to achieve high accuracy and efficiency in CNNs. `EfficientNet`, proposed by Tan & Le (2019), innovatively addresses this challenge through the implementation of compound scaling and utilization of Mobile Inverted Residual Bottleneck Convolution (MBConv) blocks (Sandler et al. 2018). This effective strategy yields state-of-the-art performance across numerous tasks and data sets. While the paradigm of CNNs continues to evolve and transformer-based models like the Vision Transformer (ViT) model (Dosovitskiy et al. 2020) have also shown excellent performance in image classification tasks, models based on the `EfficientNet` framework are still among the top-ranked models on mainstream data sets (Tan & Le 2021) such as ImageNet (Deng et al. 2009), CIFAR-10, CIFAR-100 (Krizhevsky 2009), and ImageNet21k. Additionally, VIT requires a significant amount of self-attention computation, resulting in greater computational resources and time during training and inference. Taking into account both model performance and computational complexity, we used `EfficientNet-B3` architecture as our backbone but included several modifications.

We implemented two fully connected blocks as the header of the model, and we included two additional batch normalization layers after the pooling layer and before the final output, respectively. Batch normalization is a useful regularization method to adjust the distribution of feature maps and smooth losses by calculating the mean and variance of the input features, thereby improving the performance of the model (Ioffe & Szegedy 2015). For the final output layer, we specified the `softmax` function to efficiently produce normalized probabilities of each class. The structure of the model architecture is illustrated at the top of Figure 3. The input layer takes the raw image and feeds it into a series of MBConv blocks, which are responsible for extracting relevant features. This critical architectural component of `EfficientNet`, depicted at the bottom of Figure 3, is specifically designed to enhance the representation of the network. Each MBConv block first decomposes standard convolutions into a point-wise convolution and a depth-wise convolution. This innovative approach substantially reduces the computational demands and the model parameters while preserving a robust representational capacity. Following this, the squeeze-and-excitation mechanism is applied to recalibrate the feature maps, enabling the network to focus more effectively on informative features. Finally, the Projection Layer and Residual Connection work together to merge the transformed and original features, thus maintaining representational capacity. The orchestrated operations within each MBConv block not only enhance performance but also promote efficiency. In order to highlight the advantages of the `EfficientNet` model built on MBConv blocks, we conducted a comparative study with `ResNet`, a renowned model in the field of deep learning for image classification. `ResNet`, constructed based on residual blocks, has seen wide adoption owing to its robust yet simple design and its exceptional performance on various image classification tasks, establishing it as a standard baseline in the field. Given this, we utilized `ResNet-34` as our model's backbone for the comparative experiments. The results of this comparative analysis are detailed in Section 3.1.

To train the model, it is necessary to specify a suitable error function that minimizes the difference between the true value and the predicted value for each input image (Hastie et al. 2009). We defined the `cross-entropy` loss with the `label smoothing` method (Szegedy et al. 2016) as our loss function. The `cross-entropy` loss is defined as

$$\mathcal{L} = -\frac{1}{N}\left[\sum_{i=1}^{N} y_i \log(p_i) + (1 - y_i)\log(1 - p_i)\right], \quad (2)$$

where $y_i$ and $p_i$ are label vector and softmax probability of sample *i*, respectively. *N* is the size of the batches. The original $y_i$ is a one-hot encoded vector taking a value of either 1 or 0, representing binary and single-star labels, respectively. This encoding typically leads to the widest possible gap between the logits of each class, thus causing the model to become overconfident in its label predictions. Instead of using the one-hot encoded label vector, `label smoothing` introduces noise by soft label assignment, where the labels are mixed with the uniform distribution. Parameter $y_i$ is then given by

$$y_i = \begin{cases} 1 - \epsilon, & \text{if } i \in C, \\ \frac{\epsilon}{K-1} & \text{otherwise}, \end{cases} \quad (3)$$

where *C* is the correct class and *K* is the number of classes. Parameter $\epsilon$ is the label smoothing factor, set to 0.1 by default. `Label smoothing` is a regularization method that prevents the model from overfitting and poor generalization, thus improving the performance of the model (Müller et al. 2019).

For performance metrics, we used accuracy and area under the curve (AUC) to assess the reliability and validity of the model. The accuracy is the ratio of correctly classified samples to total samples. The AUC is defined as the area under the receiver operating characteristic (ROC) curve, which calculates the true positive rates (TPR) and the false positive rates (FPR) using different probability thresholds (Bradley 1997). TPR and FPR are given by

$$\text{TPR} = \frac{\text{TP}}{\text{TP} + \text{FN}} \quad (4)$$

$$\text{FPR} = \frac{\text{FP}}{\text{FP} + \text{TN}}, \quad (5)$$

where true positives (TP) are correctly classified binaries and false negatives (FN) are the ones that are incorrectly classified as binaries. False positives (FP) denote the single stars classified as binaries. Correctly classified single stars are true negatives (TN).

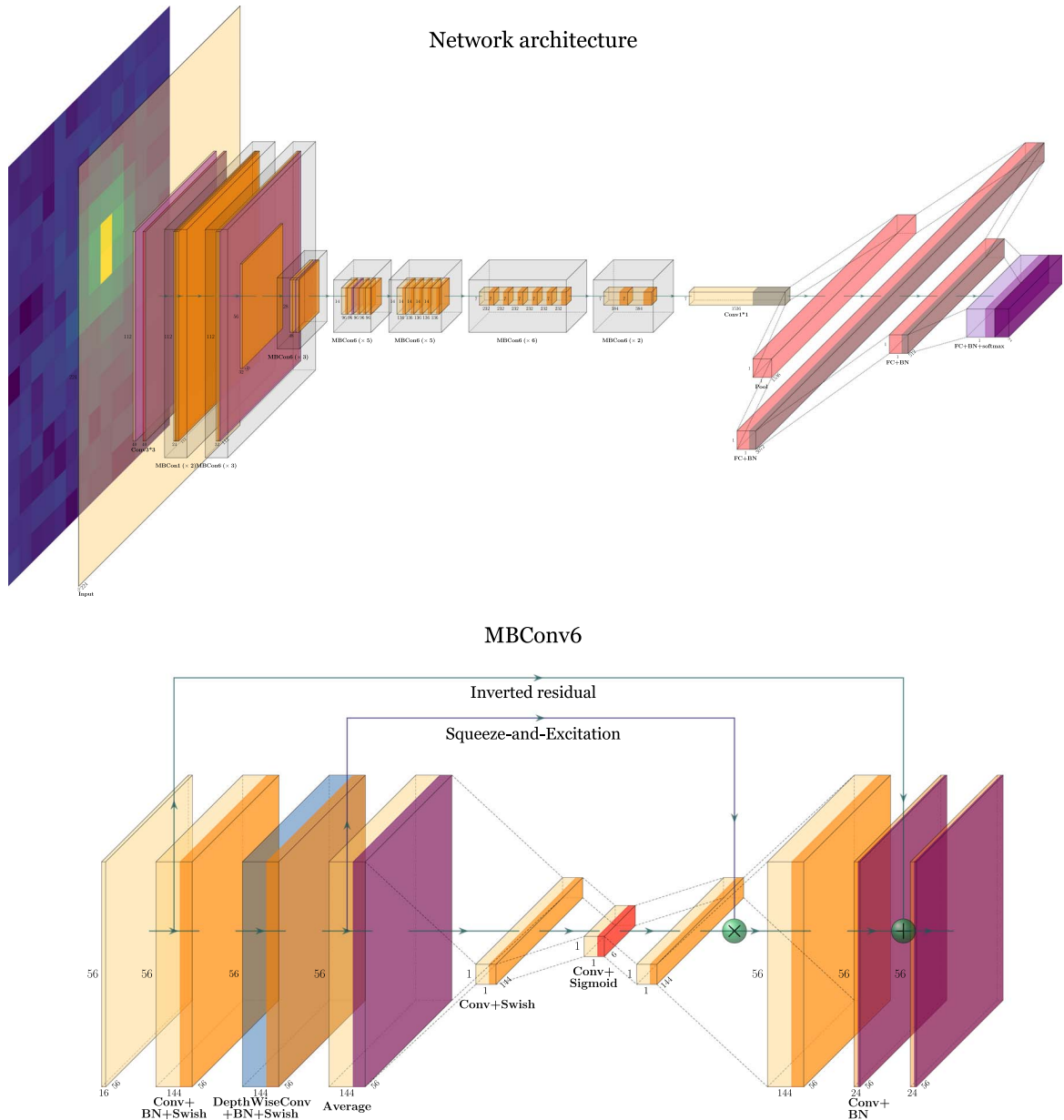

**Figure 3.** Top panel: the framework for binary detection. The network architecture uses `EfficientNet-B3` as the backbone, supplemented with fully connected layers and batch normalization layers. Bottom panel: illustration of the MBconv6 block in the `EfficientNet-B3` model, which is composed of three main components: depth-wise separable convolution, expansion convolution, and squeeze-and-excitation module.

A great feature of the ROC curve is that it remains consistent even when the distribution of positive and negative samples in the test set changes (Fawcett 2006). The AUC scores generally range between 0.5 and 1. AUC = 0.5 means that the model performs identically to a random classifier, while the model is able to correctly distinguish all positive and negative classes when AUC = 1. The closer the AUC score is to 1 and the closer the ROC curve is to the upper left corner, the better the performance of the model.

### 2.5. Training Process

In the training phase, we divided our data sets into two parts: 80% of the sample as the training set and 20% reserved for validation; we also maintained a balance of the number of single stars and binaries in both training and validation sets. The implementation of the deep-learning model in this paper is done through the deep-learning library of *PyTorch 1.9.0* (Paszke et al. 2019).

In practice, we trained our model in batches of 64 images, and the entire sample of the training set was passed through the network once as one epoch. The initial learning rate is set to 0.004, and the model is trained to minimize the loss function using a gradient-based optimizer called the `Adam` (Kingma & Ba 2014) optimizer. To train our model efficiently, we performed a learning rate schedule called `One cycle policy` (Smith & Topin 2019), which contains two learning rate steps: one is to increase the learning rate, and the other reduces the learning rate. After this, the learning rate decreases further over the iterations, several orders of magnitude below the initial value. The `One cycle policy` helps the model converge to the global optimum quickly and shorten the training time. The training process[11] is repeated for several epochs until the loss of the model converges in both the training and validation sets with no further improvement in precision. This process was executed on four Nvidia Tesla P100 GPUs, requiring approximately 50 GB of GPU memory and around 180 million FLOPs in computational overhead.

## 3. Results

### 3.1. Model Performance

As we described in Section 2.5, we retained 20% of the data set for validation, that is, 14,568 images, about half of which are single stars and half are binaries. After 30 epochs of training with the model based on the `ResNet-34` backbone, the accuracy converged to 86%, with an AUC score of 0.944. When using our model, which is based on the `Efficient-Net-B3` backbone leveraging MBConv blocks, the model achieved an accuracy of 97.2% and an AUC score of 0.997 after 30 epochs of training. This represents an increase of 11.2% in accuracy and a significant improvement in the AUC score when compared with the results from the model based on the `ResNet-34` backbone, demonstrating the superior performance of our proposed model for this task. The corresponding ROC curve and confusion matrix of our model based on the `EfficientNet-B3` backbone are displayed in Figure 4. Our model reached a high-performance level for mock images of CSST, as demonstrated by the ROC curve, which lies close to the upper left corner. Almost all samples in the validation set are correctly classified. Specifically, only 2.3% of single stars are erroneously classified, and 3.2% of binaries are mistaken for single stars. This also implies that the capabilities of the model to distinguish between single stars and binaries appear to be balanced.

To understand how data features affect model predictions, we applied a feature attribution technique known as `Occlusion` (Zeiler & Fergus 2014) to determine whether the model truly identifies the binaries in the image. The important feature regions that influence the probability scores of the classification can be visualized using the `Occlusion` technique. By means of occluding different regions in the original image, the `Occlusion` technique quantifies the attribution on the model decision with a given stride and sliding window (Ancona et al. 2018). It is an iterative process where each region is assigned a value that can be interpreted as the importance score for prediction until all regions of the image are completely covered.

---

[11] Our code is available at https://github.com/seasuf/Unresolved-Wide-Binaries.git, while the sample images utilized can be accessed at https://deepSolar.quickconnect.cn/sharing/PM4p3ccam.

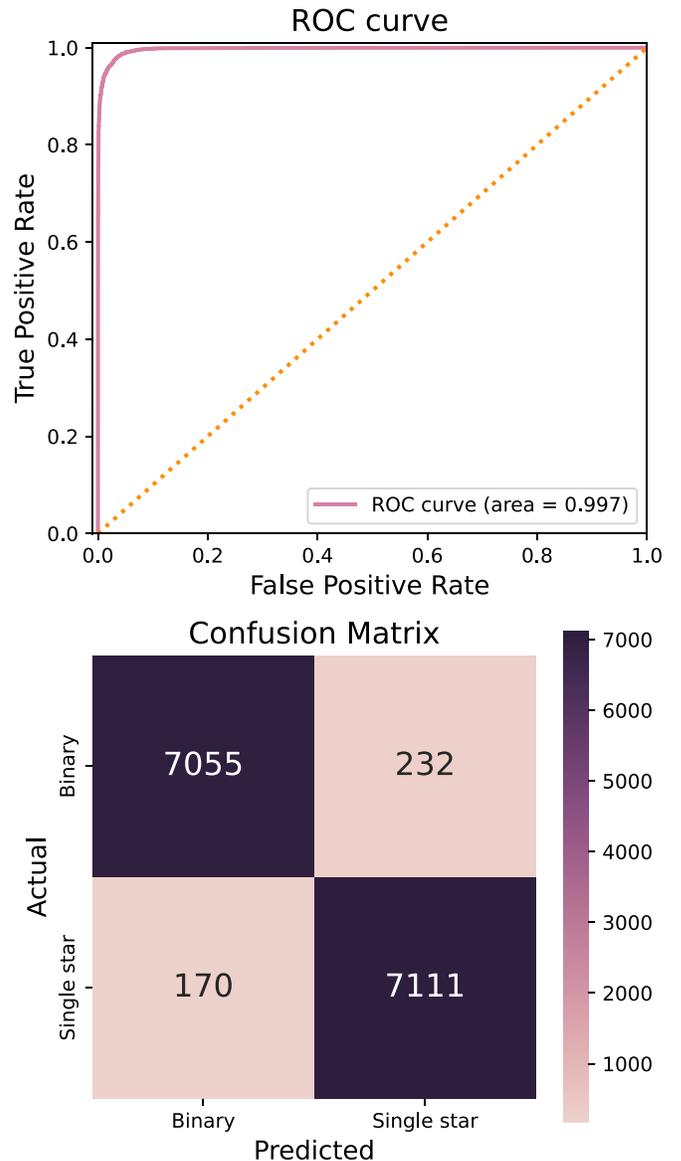

**Figure 4.** Top panel: ROC curve of the model based on the `EfficientNet-B3` backbone for CSST data. The yellow dotted line represents the random classifier, and AUC is derived by calculating the area under the ROC curve (pink line). Bottom panel: confusion matrix of the model. The *x*-axis and the *y*-axis are the predicted label and true label, respectively. The color bar and value in each box correspond to the number of the sample.

We employed the *Captum* package (Kokhlikyan et al. 2020) to investigate model interpretability, with the results of this analysis being presented in Figure 5, where the left panel shows a binary with a separation of 1 pixel and the corresponding occlusion-based attribution map is exhibited in the right panel. The left panel of Figure 5 represents a 14 × 14 physical pixel region on the CCD, displayed at a 224 × 224 pixel resolution. The right panel shows the occlusion-based attribution map at the same 224 × 224 resolution. The attribution map distinctly highlights the image regions that contribute to the model's classification decision, using varying colors to represent the importance score of each region. The color bar reflects the degree of importance associated with individual regions, with lighter shades signifying greater relevance of the contained features for making accurate predictions. As can be observed in Figure 5, the regions with high importance scores in the attribute map are concentrated at the locations of the binary in the original image, meaning that the regions most relevant to the prediction are mainly in pixels occupied by the binary, which is also consistent with our expectation. Furthermore, the model is able to identify high-frequency information such as the edge or center of the binary. This visualization intuitively interprets the intricate decision-making process of the model and assesses the impact of each pixel on the model's output.

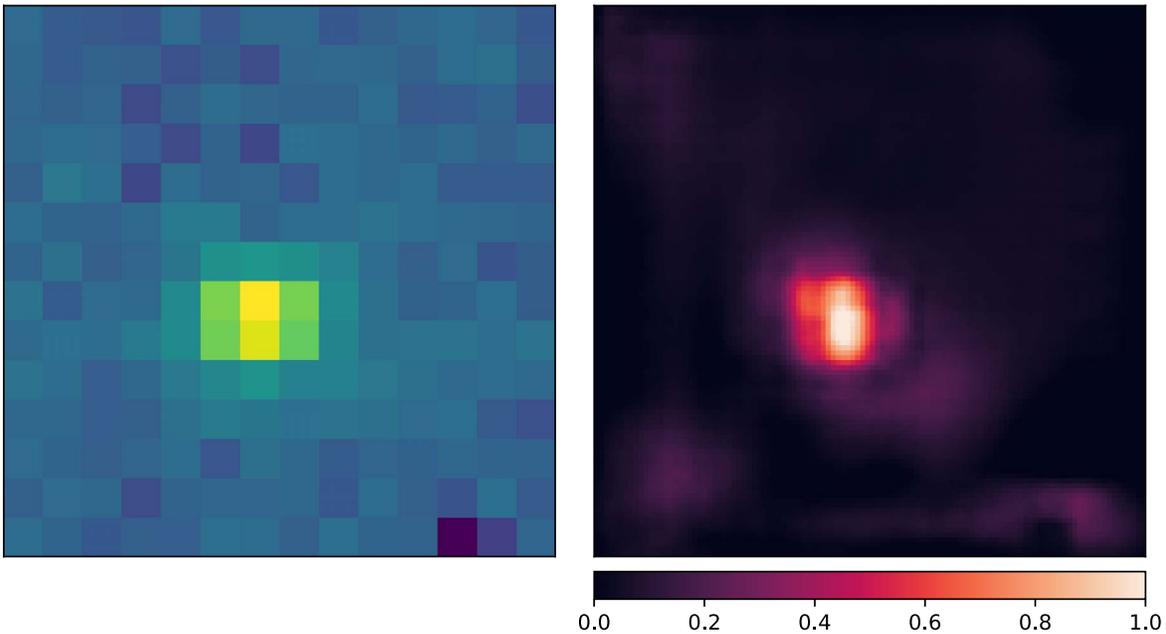

**Figure 5.** Left panel: example of a binary with 1.0 physical pixel separation. Right panel: the occlusion-based attribution map is employed to visualize the features that our model relies on for classification tasks. Features that make a significant positive contribution to the model's decision are denoted by lighter shades, while those that negatively impact the model's decision are illustrated with darker shades.

### 3.2. Impact of Noise

S/N is a crucial factor that influences the performance of a network. In this study, we conduct an experiment to systematically examine the relationship between the accuracy of the network and the S/N of the input images. As described in Section 2.2, the backgrounds of the mock images comprise Gaussian and Poisson noise, while the S/N can be controlled by assigning a specific flux to each star. Utilizing Equation (1), we generated six supplementary test sets containing mock images with distinct S/Ns, specifically S/N = 30, 55, 80, 105, 130, 155. Each test set encompasses 6000 images, evenly divided between single stars and binaries. Subsequently, the trained model is employed to predict the labels of the images in each of these test sets, and the accuracy of the model is computed, as illustrated in Figure 6. By testing our model on these supplementary test sets, we can better understand its performance under a variety of observational conditions and demonstrate its robustness and generalizability.

We see that the accuracy is about 80% at the lowest S/N (30), which rapidly increases to above 93% when the S/N is greater than 55. As for a very high S/N (155), the accuracy of the model can approach 99%. This is attributed to the fact that the capacity of the model to capture image features is significantly enhanced as the S/N increases. Our model is robust and maintains a high level of performance even at a very low S/N, and the benefit of its increase is marginal to accuracy when the S/N exceeds 80.

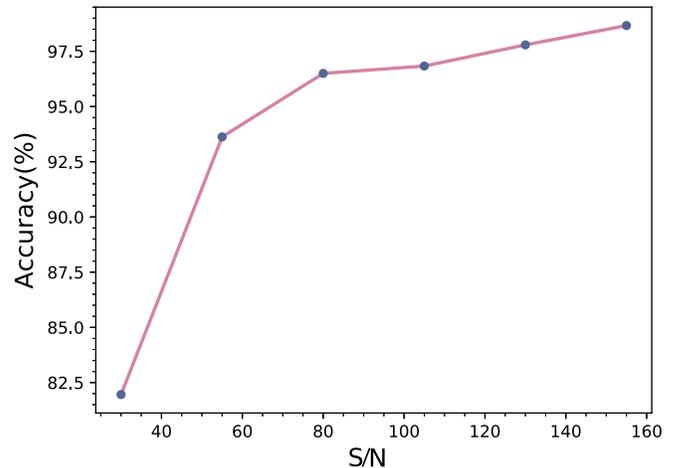

**Figure 6.** Accuracy as a function of S/N. The pink line links the circles, which reflect the accuracy at various S/Ns.

### 3.3. Impact of Binary Separation

Our model is intended to detect binaries with a separation within 0.1–2 physical pixels, so the sensitivity of the model to binaries with varying separations is necessary to explore. Following the prescription in Section 2.2, we also generated an additional batch of test sets to evaluate the performance of the model for binaries with varying separations. These test sets contain binaries with separations from 0.1 to 2 physical pixels (with step of 0.1 pixels), respectively, and each test set consists of 3000 images that maintain the same uniform distribution of S/N as the training set. We then infer those test sets through the trained model, and the accuracy of each test set is shown in Figure 7.

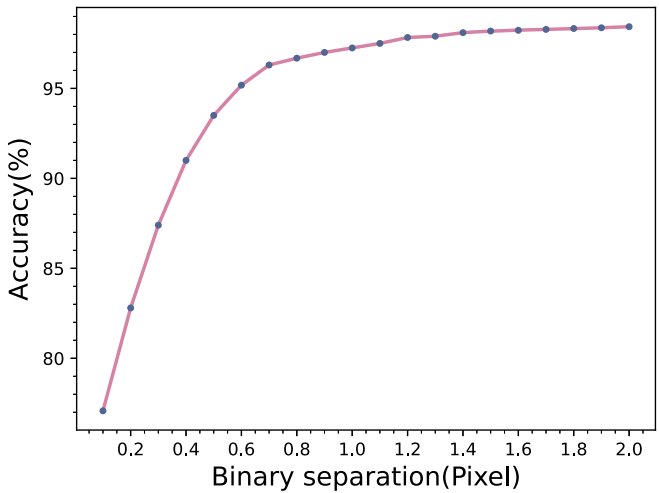

**Figure 7.** Accuracy as a function of binary separations. The circles represent the accuracy of the separation in 0.1 physical pixel steps, from 0.1 to 1 physical pixel, and are connected by pink lines.

The accuracy of our model clearly declines as the binary separation decreases, with a noticeable drop when the separation is less than 0.5 pixels. However, the accuracy remains at 77% as the binary separation approaches 0.1 pixels. In contrast, the accuracy tends to plateau for separations greater than 1.2 pixels, achieving its peak accuracy of 98.4% when nearing 2 pixels. The observed trend in our model's performance across various binary separations aligns with expectations, as the distinguishing features of binary stars become more apparent and easier to detect when they are farther apart. However, as the separation between binary stars narrows, their observable features increasingly overlap and become indistinguishable from those of single stars, making their identification more difficult. As a result, relying solely on image-based methods for detecting unresolved binaries with separations of 0.1 physical pixels or less may not produce reliable outcomes.

## 4. Validation

To verify our method, we now apply our method to HST WFC3, starting by generating mock images of HST to train the model and then performing the well-trained model to detect binaries in NGC 6121.

### 4.1. Photometry on NGC 6121

Clusters are an ideal environment to validate our method for detecting wide binaries, and the surviving binaries in the dense environment also reveal the origin and evolution of clusters. As previously mentioned, the components of binaries may overlap in the observed images when they are close enough to be indistinguishable through telescopic observation. In such cases, these binaries will be treated as single point sources in photometry, leading to the formation of binary sequences. Binary sequences are evident in the CMD of many clusters, where they are brighter than the single MS (e.g., Bellazzini et al. 2002; Ivanova et al. 2005; Hu et al. 2010). Due to the superposition of two components, the magnitude of these

binaries can be estimated by

$$m_b = m_1 - 2.5 \log_{10}\left(1 + \frac{F_2}{F_1}\right), \quad (6)$$

where $m_1$ is the magnitude of one component and $m_b$ is the magnitude of an unresolved binary. $F_1$ and $F_2$ are the flux of each component, which are positively correlated with mass in the case of MS–MS binary stars. When they are equal-mass binaries, the $m_b$ will reach a maximum of 0.75 mag brighter than a single star. Otherwise, they lie mainly between the single MS and equal-mass binary lines in the CMD. Thus, it is expected that by our method wide binaries would be more easily identified in the binary sequences, and we can also estimate their binary separations if the distance of the clusters is known.

NGC 6121 (M4) is a globular cluster with an age of 12.7 Gyr (see Hansen et al. 2002) and the closest globular cluster to Earth, with a distance of 1.72 kpc (Peterson et al. 1995). It has been well studied with abundant observation data (e.g., Richer et al. 2004; Marino et al. 2008). It is noteworthy that although NGC 6121 appears to have surpassed its dynamical relaxation time, there is no evidence of a central brightness cusp (Trager et al. 1995), implying that NGC 6121 has not entered the phase of core collapse. This phenomenon may be attributed to the interaction of the central binaries preventing the core collapse (Cote & Fischer 1996). Milone et al. (2012) have confirmed that a relatively large proportion of stars in NGC 6121 are binaries (approximately 10%–15%). Consequently, this cluster serves as an ideal subject for our study.

To derive the photometry on NGC 6121, we selected the F467M and F775W filters of the ultraviolet and visual (UVIS) channel of the Wide Field Camera 3 (WFC3) instrument on board HST, using data from the program GO-12911: "A search for binaries with massive companions in the core of the closest globular cluster M4" (PI: Bedin). The description of program GO-12911 is summarized in Bedin et al. (2013). These two filters are chosen because of their relatively stable PSF and better astrometric characteristics. The WFC3/UVIS has two 2051 × 4096 pixel CCDs with a field of view of 160″ × 160″ and a very high spatial resolution (PSF FWHM ∼ 0.″08). Our analysis is performed on individual flat-fielded images ("flt" type), as these are obtained from direct observations, which is necessary for subsequent PSF study, whereas the resampled data after the Drizzle procedure are unsuitable for high-precision analysis of the PSF.

We first used the DAOFIND (Stetson 1987) algorithm to detect star sources with the DAOFIND MMM routine for background estimation in the field. Then, we performed crowded-field photometry using the Photutils package (v1.2.0; Bradley et al. 2021) provided by Astropy (Astropy Collaboration et al. 2013, 2018). Since the goal of this process is to isolate the binary-sequence stars from the single MS stars in the CMD for subsequent model inference, it is not necessary to independently calibrate the magnitude of each star. Finally, the instrumental CMD is derived after matching the stellar coordinates of each filter in a single field.

Figure 8 compared the instrumental magnitude $m_{F775W}$ versus $m_{F467M} - m_{F775W}$ CMD obtained from the photometry. We can clearly see that the stars are clustered in two sequences (left panel), a single MS at a lower magnitude and a broadening of about 0.75 mag above the single MS, which is the binary sequence. As a validation, we removed uncertain sources with fainter magnitudes and selected a segment of the single MS and binary sequence, from instrumental magnitude $m_{F775W} \sim -9$ up to $m_{F775W} \sim -11$, for applying our method to identify wide binaries and comparing their proportions between the binary sequence and single MS. The single MS and binary-sequence regions of concern to us are shown in the right panel of Figure 8, containing 1848 and 102 sources, respectively.

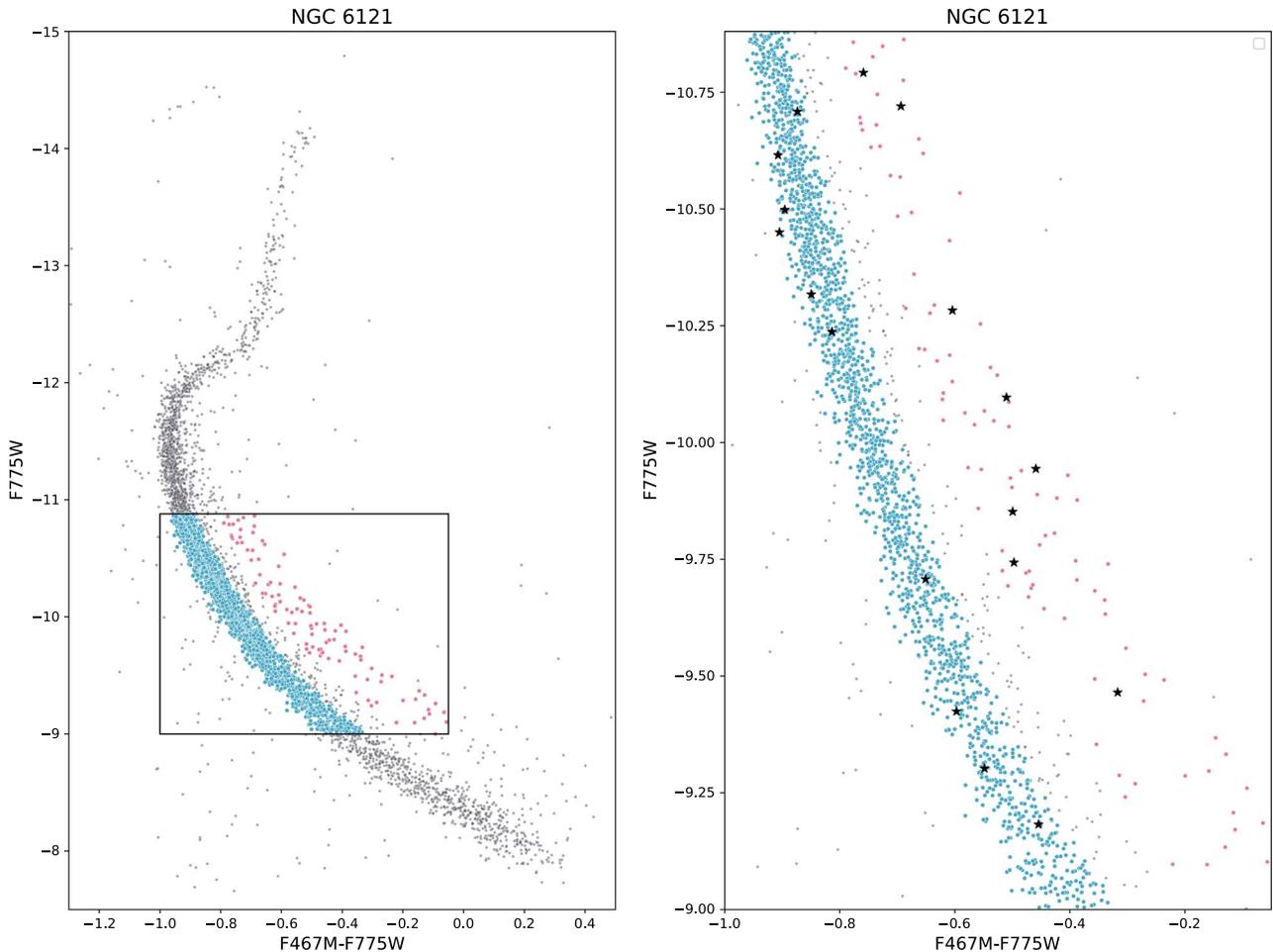

**Figure 8.** Left panel: CMD of NGC 6121 using instrumental magnitude $m_{F775W}$ vs. $m_{F467M} - m_{F775W}$. Red circles are objects tagged as MS stars, and cyan circles are objects tagged as binary-sequence stars; Right panel: same as the left panel, but zooming in around the MS and binary-sequence regions. The wide binary candidates are indicated by the black star symbols (see Section 4.3).

### 4.2. Data Set Generation for HST

In this study, we used images from the F775W band for our validation, necessitating the generation of training data for this band. To simulate the mock images of WFC3/UVIS, we constructed the PSF models using the effective PSF (ePSF) technique developed by Anderson & King (2000) and Anderson (2016). This approach allows us to reproduce real stars as accurately as possible. The ePSF is a smooth continuous function extracted from the grid points of each star. Due to the undersampling of WFC3/UVIS, the grid points can be obtained by a ×4 oversampling of the image pixels. To accomplish this, we selected 25 isolated stars in the single MS regions of the CMD of NGC 6121, which have good-quality profiles with clean backgrounds. Each star is cut to 25 × 25 pixels in the F775W band image. We then constructed the ePSF model as a 101 × 101 grid by performing the EPSFBuilder procedure (part of the Photutils software package).

After extracting the ePSF model in the F775W band, we generated the mock images of HST by convolving the point source with the ePSF, and we cropped the images to a size of 14 × 14 pixels. Considering the instrumental PSF FWHM (approximately 0″.08) and the physical pixel size (approximately 0″.04) of WFC3/UVIS, we assumed that any two stars within 2 physical pixels of each other are considered to be unresolved binaries. Essentially, the generation details are the same as those performed for the CSST data set (Section 2.2). The separation of binaries ranges from 0.1 to 2 physical pixels evenly, and the fluxes of both single stars and binaries are uniformly distributed between 5000 and 25,000. Note that we estimated the background level from the real HST images of the F775W band.

Finally, we obtained about 54,880 mock images of single stars and the same number of binaries. The constructed ePSF model in the F775W band and examples of mock images can be seen in Figure 9, i.e., the ePSF model of F775W (panel (a)), a single star (panel (b)), a binary with 0.5-pixel separation (panel (c)), and a binary with 1-pixel separation (panel (d)). We also kept the background level and total fluxes consistent in panels (b), (c), and (d). The ePSF model consists of a 101 × 101 metric, which has a lower concentration of energy than the PSF of CSST, so that the stellar profiles are significantly larger than those of the CSST images.

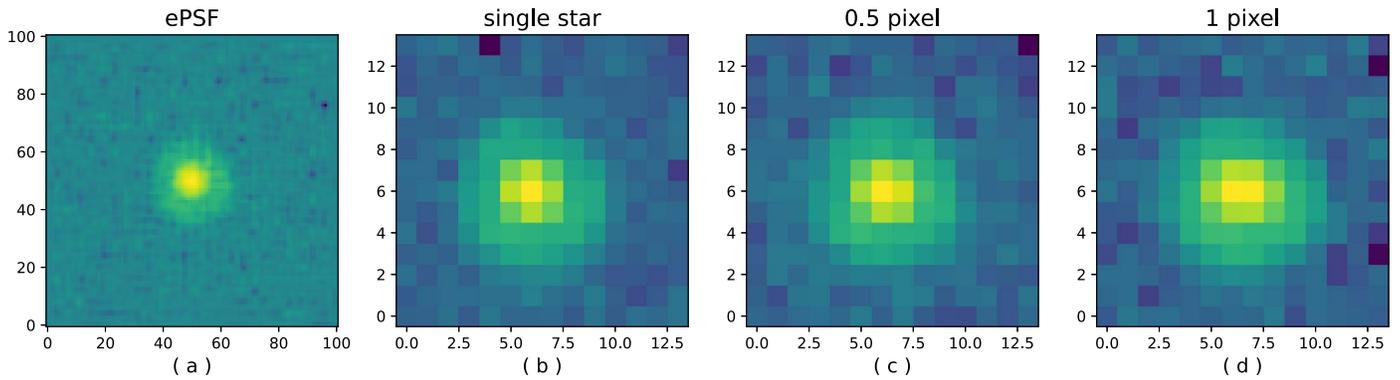

Figure 9. Examples of the ePSF and simulated images for the F775W band of HST WFC3/UVIS. (a) A log image of the ePSF model in the F775W band. (b) Log image of a single star (flux of star $F = 10,000$). (c) Log image of a binary with 0.5 physical pixel separation ($F_1 = F_2 = 5000$). (d) Log image of a binary with 1 physical pixel separation ($F_1 = F_2 = 5000$).

### 4.3. Detecting Wide Binaries in NGC 6121

In this section, we intend to identify wide binaries in NGC 6121 based on the deep-learning approach. The network is trained on the mock images of HST, with 20% of the data retained for validation. The training process follows that in Section 2; we repeated the entire procedure, including the data augmentation, network structure, loss functions, training strategies, etc. We trained the model for 35 epochs and achieve an accuracy of 96.5%, which is slightly lower compared to the model trained on CSST. The ROC curve and confusion matrix are presented in Figure 10. the AUC score is 0.996, and the ROC curve reaches the upper left corner of the plot, indicating the good performance of the model. The confusion matrix shows that the model is not biased toward recognizing a particular class, with approximately 3.2% of the single stars and 3.5% of the binaries being erroneously identified.

Consequently, the model appears to be highly confident in the prediction that it can be used for practical applications. We first tagged all the sources within the red and blue regions in Figure 8 onto the F775W band image of NGC 6121 and then applied the well-trained model to infer the images of these stars for predicting whether they are wide binaries. Ultimately, our model detected 8 wide binaries out of a total of 102 stars in the red regions of Figure 8, as well as 10 wide binaries out of a total of 1848 stars in the blue regions. These wide binary candidates are marked with black star symbols in the right panel of Figure 8; we can see that the wide binary candidates in the binary sequence are more pronounced and appear to be distributed at the edges of the binary sequence. Taking the uncertainties into account, the proportion of wide binaries in the binary sequence is approximately $7.84\% \pm 2.66\%$, in contrast to the $0.54\% \pm 0.17\%$ observed in the single MS. Since the capability of our model is to detect binaries within 0.1–2 physical pixels apart, which translates into an angular separation of $0\rlap{.}''004$–$0\rlap{.}''08$ for WFC3/UVIS, the physical separation of these binaries can therefore be estimated to be around 7–140 au in light of the distance of 1.72 kpc for NGC 6121 (Peterson et al. 1995).

For the contamination of chance alignments, we calculated the stellar density in the HST field of view by counting the sources of NGC 6121 in the F775W band, and then we implemented the Monte Carlo simulations as in Section 2.1.

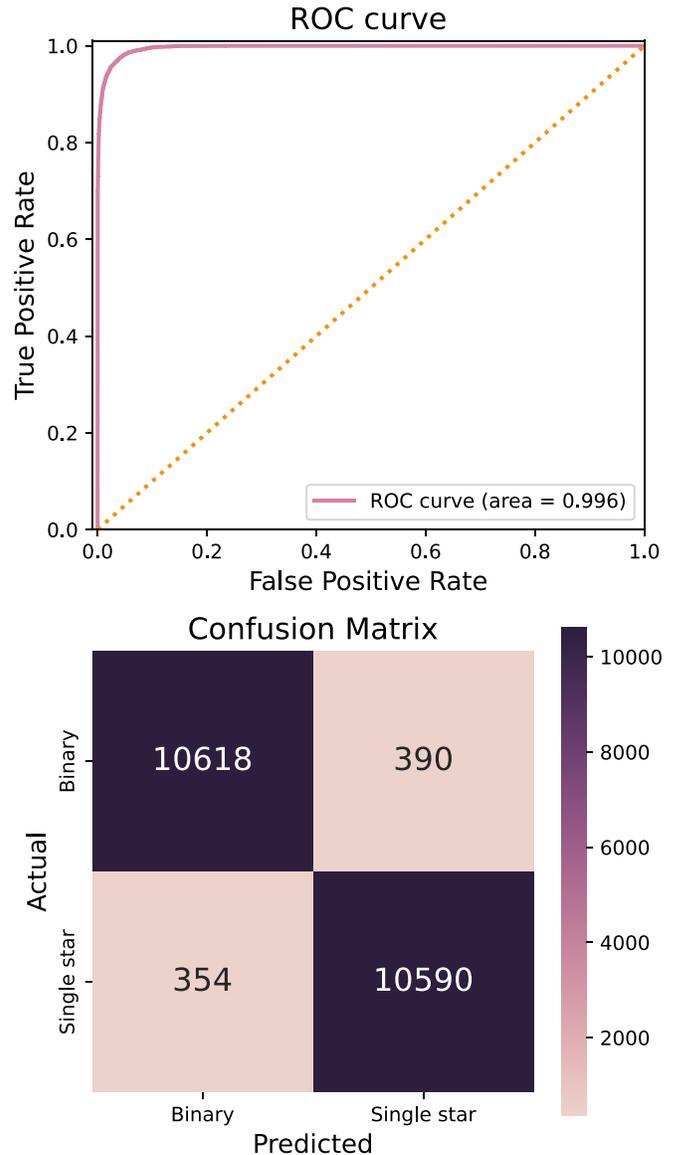

Figure 10. Similar to Figure 4, but for the F775W band of HST WFC3/UVIS. Top: ROC curve of the model; bottom: confusion matrix of the validation set.

We estimated the contamination of chance alignments to be about 0.3%, where a chance alignment here is defined as any two stars with an angular separation of less than $0\rlap{.}''08$. Among the detected wide binary candidates, eight are located within the binary sequence. Considering the contamination of chance alignments, the number of chance alignments in the binary sequence is estimated to be $0.3 \pm 0.6$.[12] This low probability of chance alignments in the binary sequence allows us to confidently assert that the eight candidates are not chance alignments. In contrast, 10 wide binary candidates are found within the single MS, where the number of chance alignments is estimated to be $5.6 \pm 2.4$. While the number of these candidates exceeds the expected quantity for chance alignments, it is still consistent with a $3\sigma$ error. These candidates within the single MS could potentially represent wide binaries with large luminosity ratio disparities, making them difficult to distinguish from single stars. Nevertheless, we cannot completely exclude the possibility that they are chance alignments. This observation supports the theoretical expectation that wide binaries should be more frequently encountered in the binary sequence. The presence of a higher proportion of wide binary candidates within the binary sequence, coupled with the low probability of chance alignments in this region, provides indirect evidence of the effectiveness of our method in detecting unresolved wide binaries. This suggests that our approach could be successful in identifying unresolved wide binaries. However, further investigation and validation are necessary to confirm these findings.

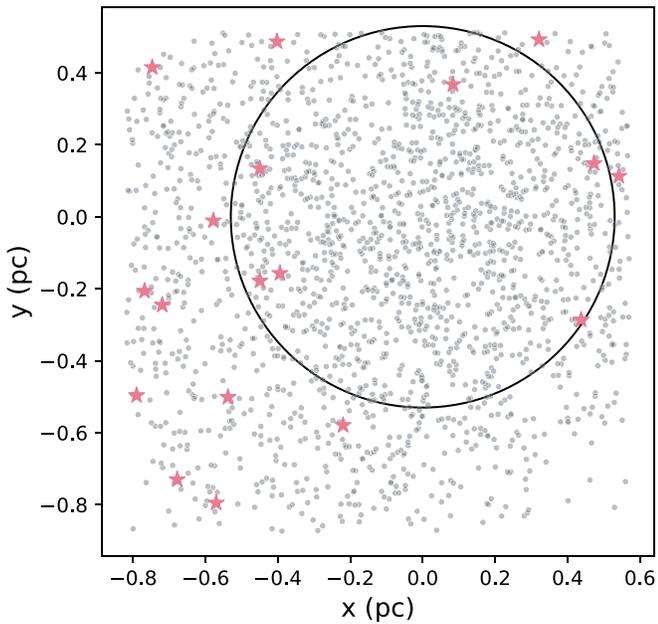

**Figure 11.** Gray circles show the spatial distribution of all sources in the regions of our concern in NGC 6121, and the red star symbols represent the wide binary candidates. The black circle indicates the cluster core of NGC 6121 (radius ∼ 0.53 pc).

### 4.4. Statistical Assessment of the Results

To have a further validation of our results, we calculated the positions of all the sources we obtained relative to the cluster center. Figure 11 presents the spatial distribution of all sources. The core radius area of NGC 6121 is about 0.53 pc (Trager et al. 1993), indicated by the black circle. We found that the fraction of wide binary candidates is $0.49\% \pm 0.2\%$ inside the cluster core and $1.65\% \pm 0.47\%$ outside the core radius. A statistical test (Z-test) reveals a significant difference between these proportions ($Z = 2.58$, $P = 0.00989$), indicating that the proportion of wide binaries (7 au $< a <$ 140 au) in the core of NGC 6121 may be lower than the proportion outside the core. Considering the binary proportion reported by Milone et al. (2012), it is possible that the majority of binaries observed in the core of NGC 6121 consist of close binaries with smaller separations. It is important to note that this conclusion is subject to the limitations and constraints of our study, particularly the incompleteness of our wide binary candidate sample, which is primarily due to the parameter space limitations, such as the magnitude range and the separation range covered in our study. We focused on separations from 0.1 to 2 physical pixels (corresponding to 7–140 au) and a specific magnitude range owing to the constraints of our deep-learning method. Therefore, our results do not encompass the complete binary population, especially those with separations beyond our studied range or outside our magnitude limitations. By acknowledging these limitations, we emphasize that our findings should be interpreted with caution, and future research may shed more light on the true proportion of wide binaries in NGC 6121.

In the top panel of Figure 12, we present the cumulative distribution function (CDF) of the distance from the center of NGC 6121 for wide binary candidates and all sources. The CDF of wide binary candidates (pink line) appears shifted to the right of the CDF of all sources (gray line), indicating that the former sample tends to have larger distances from the center of the cluster. To evaluate the statistical significance of this difference, we employed a two-sample Kolmogorov–Smirnov test to assess the significance of the differences in their respective distributions. The result was a D statistic of 0.41 and a *p*-value of 0.3%, where the D statistic measures the maximum difference between the two CDFs and the probability represents the likelihood of obtaining such a difference by chance. The observed large value of the D statistic and small *p*-value provide compelling evidence to reject the null hypothesis of identical underlying distributions between the two samples.

To further visualize the difference between the two samples, we created a quantile–quantile plot (Q–Q plot) in the bottom panel of Figure 12, where the quantiles of the two samples are visualized against each other. If the samples correspond to the same probability distribution, the points should be along the 45° line (yellow dotted line). Obviously, the Q–Q plot shows that the quantiles of wide binary candidates are quite different from the quantiles of all sources, indicating that the two samples come from different distributions. This difference in distributions can indicate that wide binary candidates may have a different origin or formation history than other sources in NGC 6121, and further analysis and investigation may be needed to understand the underlying reasons for this difference.

Many studies of multiple populations in clusters reveal the existence of primordial differences between first-generation (FG) and second-generation (SG) stars (e.g., Carretta et al. 2009; Milone et al. 2017; Tailo et al. 2019). As both observational and theoretical studies have shown, SG stars form in the central regions of clusters and are concentrated in the dense environments during their evolution, while FG stars are mainly distributed in the sparse environments (e.g., Sollima et al. 2007b; D'Ercole et al. 2008; Richer et al. 2013). Vesperini et al. (2011) and Hong et al. (2015, 2016) investigated the evolution of binaries in clusters by means of N-body simulations, and the results showed that the disruption rate of SG binaries is significantly higher than that of the FG binaries.

---

[12] Please note that in cases where the expected counts are small, as is the case here, the corresponding standard deviation can indeed be large.

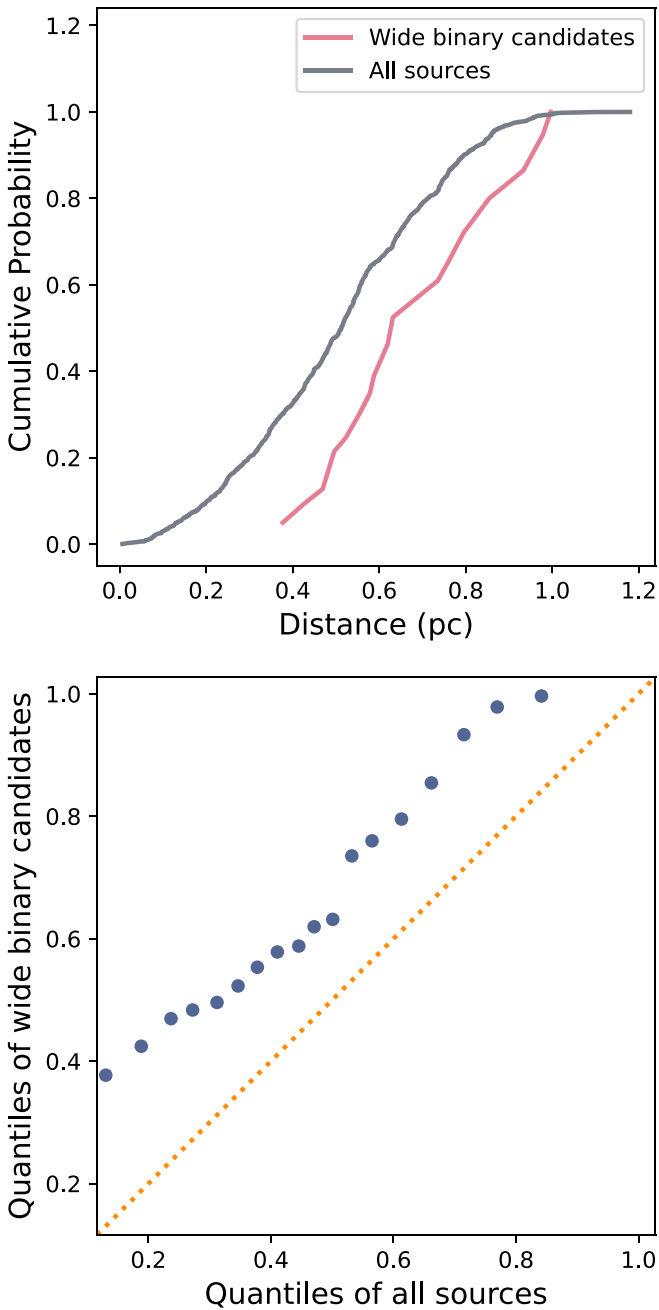

**Figure 12.** Top panel: CDF of the distance from the center of NGC 6121 for all sources (gray line) and wide binary candidates (pink line). Bottom panel: the corresponding quantile–quantile plot, with a 45° line (yellow dotted line) plotted for reference.

Environment conditions play a crucial role in the survival and distribution of wide binaries in clusters (Hong et al. 2019). FG binaries, which are born in more relaxed and sparse environments, are more likely to be found outside the core radius, allowing wide binaries to survive in larger numbers. This contrasts with the more concentrated SG population that experiences a higher disruption rate owing to dynamical processes in the dense cluster center (Calura et al. 2019). The latest Monte Carlo simulations of binaries in globular clusters performed by Sollima et al. (2022) showed that wide SG binaries ($a > 10$ au) are immediately destroyed during the cluster evolution, but FG binaries with a separation in the range of 10–100 au can survive even after 12 Gyr of evolution, with a binary fraction of 8%.

Although our sample of wide binaries is incomplete, considering the spatial distribution of the binaries we detected in NGC 6121 and their separations (7 au $< a <$ 140 au), it suggests that these wide binaries might primarily be FG binaries. If this is the case, our results could potentially hint at an intrinsically different binary fraction between FG and SG stars in NGC 6121, which is also supported by studies of multiple populations on NGC 6121. For instance, D'Orazi et al. (2010) found a binary fraction of 12% for FG stars and 1% for SG stars based on a study of radial velocity variations, while Milone et al. (2020) analyzed the multiple populations of binaries in NGC 6121 and showed that FG binaries dominate the binary populations. However, further research with a more comprehensive sample is needed to confirm this possibility.

It is important to emphasize that our study does not attempt to determine the binary fraction in NGC 6121, as achieving this would require a comprehensive investigation, taking into account factors such as contamination from field stars and completeness analysis. Such a detailed inquiry surpasses the scope of our current research and calls for further exploration with a more extensive sample. Despite these limitations, our findings still provide preliminary insights into the distribution and properties of such wide binaries and serve as a starting point for further research. The primary objective of our work is to demonstrate the feasibility of our approach in detecting unresolved wide binaries, which has been partially validated to some extent in observational and theoretical studies, aligning with existing findings in the field. We will continue to refine our deep-learning method and expand the parameter space to better capture the complete wide binary population in future studies. It is anticipated that the sample of wide binaries detected using our method will offer valuable contributions to the nature of multiple populations and the evolution of wide binaries within globular clusters.

## 5. Conclusion

In this paper, we have proposed a promising method to search for wide binaries that are unresolved by current observational techniques. Our method analyzes the morphology of PSF using a deep-learning technique, which takes advantage of space-based telescopes for high-resolution imaging to learn observable features of the data. We generated mock images of CSST as the training set and constructed a neural network for the purpose of identifying unresolved wide binaries, while evaluating the effect of S/N and binary separations on the results. As a validation, we explored the possibilities of applying our method to HST by analyzing photometric data of NGC 6121. Our main results are summarized below.

(1) For CSST, the training set consists of 72,840 mock images generated by the PSF in the $u$ band, which takes into account Gaussian and Poisson noises, and is divided into single stars and binaries with a separation of 0.1–2 physical pixels. We then developed a CNN network based on `EfficientNet` to identify unresolved binaries. The trained model achieves a high-accuracy

  performance of 97.2% on the mock data set of CSST, with a value of 0.997 for the AUC score.
(2) We found that the accuracy of our method for detecting binaries increased with increasing binary separation or S/N, and the accuracy significantly depends on both factors when the S/N is below 80 or the binary separation is less than 0.5 physical pixels. As a limit of detection, the accuracy is 77% for binaries with a 0.1 physical pixel separation and 80% for images with an S/N of 30.
(3) The training set of HST is generated using the ePSF, which is constructed from photometric data in the F775W band. We generated 54,880 mock images with real background noise used for the background levels. We also adopted `EfficientNet-B3` as the backbone of the model, with some modifications. The trained model is capable of identifying unresolved binaries with an accuracy of 96.5% and an AUC score of 0.996.
(4) We performed predictions for point sources on the single MS and binary sequence of NGC 6121 using HST data to validate our method. Eighteen wide binary candidates are identified out of a total of 1950 sources, with the separations ranging from 7 to 140 au. Such binaries are detected more frequently in binary sequences, while binaries in the single MS cannot be excluded as chance alignments. Their proportion is $0.49\% \pm 0.2\%$ inside the core radius of NGC 6121 and $1.65\% \pm 0.47\%$ outside the core radius. Our results are consistent with the current studies of multiple populations of clusters.

With the advent of the next generation of space-based surveys (JWST and CSST), an increasing number of high-quality data sets are becoming available. This necessitates the development of innovative methods, such as the PSF-based binary detection proposed here, to fully exploit these data. Our method holds the potential to become a powerful tool for analyzing unresolved binaries. Our approach exhibits very high accuracy in identifying binaries with separations ranging from 0.1 to 2 physical pixels, effectively addressing the challenge of dealing with previously unavailable unresolved wide binaries. In the future, our method can be employed to process space-based observations and analyze data from large surveys.

Although Gaia's spatial resolution is slightly lower than that of CSST (approximately $0\rlap{.}{''}4$), it provides a substantial number of wide binary samples through astrometry and proper-motion analysis. However, most of these samples consist of wide binaries with separations larger than 100 au, highlighting the fact that the current wide binary science based on Gaia data is somewhat limited in scope. For example, Hwang et al. (2022) used Gaia DR3 to analyze the relationship between binary separations and eccentricity distribution, but due to Gaia systematics affecting binaries beyond 100 au, their analysis excluded wide binaries with separations within $1\rlap{.}{''}5$. Moreover, the research of El-Badry et al. (2019) on twin binaries using Gaia data suggests that excess twins likely form at separations below 100 au. Therefore, obtaining additional wide binary samples with separations of less than 100 au is crucial for further investigation. Our method directly analyzes images from CSST survey data, enabling us to obtain wide binary samples with separations between tens and hundreds of au. This unique contribution that CSST can make in the field of wide binary research allows for a more comprehensive understanding of these binaries. This includes investigating the relationship between eccentricity distribution and binary separations for this range of wide binaries, exploring the formation mechanisms and evolution of excess twins, as well as comparing the wide binary fraction in star clusters, low-density star-forming regions, and the field. This research can provide valuable constraints on models for binary dynamical evolution and enhance our understanding of stellar multiplicity, as well as the environmental conditions conducive to binary formation.

In future work, we aim to refine and extend our method for validation across a broader range of clusters and more complex scenarios, with the goal of increasing the generality and accuracy of our model predictions. This involves modifying our network and expanding our training set to handle images with three or more point sources. Furthermore, we will incorporate more sophisticated and accurate PSF models. Through these enhancements, we anticipate that our model will become more robust and versatile, in preparation for processing photometric observations from larger upcoming surveys.

### Acknowledgments

We thank the anonymous referees for their constructive comments and suggestions that helped us to improve this work. We also thank Hao Tian, Xin-Ze Zhang, and Wen-Qin Sun for valuable discussion. This work is supported by the National Natural Science Foundation of China under grants (Nos. 12103064, 12125303, 12288102, 12090040/3, 12090043, 11873016), the Science Research grants from the China Manned Space Project (Nos. CMS-CSST-2021-A10, CMS-CSST-2021-A08, CMS-CSST-2021-B05), and the Joint Research Fund in Astronomy (U2031203) under cooperative agreement between the NSFC and Chinese Academy of Sciences.

### ORCID iDs

You Wu https://orcid.org/0000-0002-3616-9268
Jiao Li https://orcid.org/0000-0002-2577-1990
Chao Liu https://orcid.org/0000-0002-1802-6917
Yi Hu https://orcid.org/0000-0003-3317-4771
Long Xu https://orcid.org/0000-0002-9286-2876
Tanda Li https://orcid.org/0000-0002-5469-5149
Xuefei Chen https://orcid.org/0000-0001-5284-8001
Zhanwen Han https://orcid.org/0000-0001-9204-7778